\documentclass[runningheads]{svmult}

\usepackage{makeidx}   
\usepackage{graphicx}  
\usepackage{subeqnar}  
\usepackage{multicol}  
\usepackage{physprbb}  
\makeindex             

\newcommand{\greeksym}[1]{{\usefont{U}{psy}{m}{n}#1}}

\newcommand{\uPi}{\mbox{\greeksym{P}}}

\newcommand{\uSigma}{\mbox{\greeksym{S}}}
\newcommand{\cs}{c_{\rm s}}
\newcommand{\vA}{v_{\rm A}}
\newcommand{\vrms}{v_{\rm rms}}
\newcommand\cm{{\rm cm}}
\begin{document}
\title*{Developing Diagnostics of Molecular Clouds \protect\newline
Using Numerical MHD Simulations}
\toctitle{Developing Diagnostics of Molecular Clouds \protect\newline
From Numerical MHD Simulations  
}
%
%
\titlerunning{Cloud Diagnostics from MHD Simulations}
%
\author{Eve C. Ostriker
\authorrunning{Eve C. Ostriker}
%
%
\institute{Department of Astronomy,
University of Maryland, College Park, MD 20742, USA}
}

\maketitle              

\begin{abstract}
An important aspect of astrophysical MHD turbulence research is developing 
diagnostics to connect simulations with the observable universe.  Turbulent 
systems are by definition structurally complex in all fluid variables
(density, velocity, and magnetic field), such that they must be 
described statistically.  By developing and applying diagnostic tools to 
simulation data, it is possible to interpret empirical laws
for the statistical properties of observed systems in terms of fundamental 
dynamical processes, and to identify and calibrate robust probes of physical 
parameters that cannot be measured directly.  
Using several different 
examples, I describe how structural diagnostic analyses have already yielded 
significant insights into the nature of turbulent molecular clouds.
I review results from several different groups, and 
discuss directions for future diagnostics to enhance our understanding of 
cloud structure and constrain models of the evolutionary course that 
governs star formation.
\end{abstract}

\section{Introduction}
As the number, range, and depth of the papers in this volume
witnesses, recent progress in modeling and understanding
astrophysical MHD turbulence is impressive.  Even with the intensive
research of several groups over the past few years, however, 
many aspects of the fundamental turbulence phenomenon are not yet wholly 
understood -- which makes for continuing excitement in this emerging 
discipline.  In using the results of MHD simulations 
to interpret the dynamics of the interstellar medium, the technical challenges
involved in numerically modeling and characterizing turbulence are compounded 
by astrophysical uncertainties in posing the numerical problem to be solved.  
For molecular clouds, open astrophysical questions include:
\begin{itemize}
\item
What is the source (original, and potentially, maintaining) of turbulence?
\item
What is the mean magnetic field strength, and variation of mass-to-flux 
ratio, in molecular clouds?
\item
What is the range of sizes and masses of molecular clouds in the Milky Way and
other spiral galaxies?
\item
How are clouds formed?  how long do they live? how are they destroyed?
\end{itemize}
From the point of view of defining an idealized problem for an MHD
simulation, these astrophysical questions translate to uncertainties
in the input spectral form (in space and time) of the turbulent
driving, the value and variation of the plasma $\beta$ parameter, the
importance of self-gravity, and the initial and boundary conditions
for the simulation.

The complexity of the turbulence phenomenon demands detail and variety in 
the analytical methods used to characterize its structure.
For application to understanding astronomical 
systems -- where the physical inputs are uncertain, and only projected 
distributions are available -- the eventual aim is 
to develop a set of 
simple, robust diagnostics of MHD turbulence that have direct connections 
to observable quantities.  Potentially, there are many different avenues for
this sort of analysis, and extensive exploration is required to determine 
what directions are most productive.  
Because large-scale numerical simulations of turbulence under 
interstellar conditions are only now becoming computationally practical,
the present tasks include first
the ``forward'' process of characterizing MHD turbulence obtained from new
simulations with a range of  
parameter values, and then using these results to select and calibrate 
diagnostics for the ``reverse'' process of discriminating systemic parameters
from observables.  

As examples of the process of developing diagnostics of molecular clouds'  
internal structure, kinematics, and magnetization from turbulent MHD
simulations, my discussion here will focus on recent work analyzing
and interpreting density and column density statistics (\S 2),
properties and definitions of clumps (\S 3),
linewidth-size relations (\S 4), and statistics of polarization maps (\S 5).  
I will
update previous work (see also \cite{sto98}, \cite{ost99}, \cite{ost01}, \cite{gam01}), present several new results, and make connections to the 
conclusions of other groups.
Chapters in the same volume covering topics most directly related 
to those addressed here 
include those by Crutcher, Heiles, \& Troland; Nordlund \& Padoan; MacLow; 
Cho, Lazarian \& Vishniac; and Zweibel, Heitsch, \& Fan.

\section{Density and Column Density Statistics}

In general, a turbulent velocity field leads to production of 
significant local 
density variations in a compressible medium (i.e. a medium with 
Mach number ${\cal M}\equiv \vrms/\cs\gg 1$, 
where $\vrms$ is the turbulent velocity dispersion  
and $c_s$ the sound speed).  This is true regardless of the magnetic field 
strength, because in the case of a weak magnetic field 
($\vA\equiv B_0/\sqrt{4\pi \bar\rho} \sim c_s \ll \vrms$, where 
$\vec{B}_0$ is the mean magnetic field and $\bar\rho$ is the mean density), 
magnetic pressure forces are weak compared to ram pressure, and in the 
case of strong magnetic fields ($\vA\sim \vrms$), compression
is unhindered along the mean field direction (and indeed enhanced by 
forces associated with gradients in $B_\perp^2$, where $\vec{B}_\perp$ is the
component perpendicular to $\vec{B}_0$).  

If compression and rarefaction events are spatially and temporally 
independent, then for the case of an isothermal equation of state
(approximately true under molecular cloud conditions), the resultant one-point
density distribution function (often referred to as a ``PDF'' -- 
probability distribution function) 
is expected to obey a lognormal form (\cite{pas98}, 
\cite{nor99}).  If $1+ \delta_i$ is the enhancement/decrement factor for
density in the $i^{th}$ compression/rarefaction event affecting a given 
fluid element, then the density after $N$ events will be 
\begin{equation}
  \rho =  \bar\rho \uPi_{i=1}^N (1+\delta_i)
\end{equation} so that the logarithm of the density,
\begin{equation}
\log(\rho/\bar\rho)= \uSigma_{i=1}^N \log(1+\delta_i),
\end{equation}
will be the sum of independent random variables; by the Central Limit Theorem,
this implies that $\log(\rho/\bar\rho)$ should obey a Gaussian distribution.

The results of numerical simulations (\cite{vaz94},\cite{pad97},
\cite{pas98},\cite{ost99},\cite{ost01}) indeed bear out the expectation that 
a log-normal form for the volume density PDF prevails (at least 
away from the tails) under isothermal conditions.  This result holds both
for forced and decaying turbulence, and for simulations with varying mean
magnetic fields.  Figure 1 shows an example of the distributions of 
fractional volume and mass as a function of volume density for four 
forced-turbulence MHD simulations with ${\cal M}=5$ (see \cite{sto98} for 
details on the models), with comparisons to the lognormal functions
with the same mean and dispersion.  
\begin{figure}[t]
\begin{center}
\includegraphics[width=\textwidth]{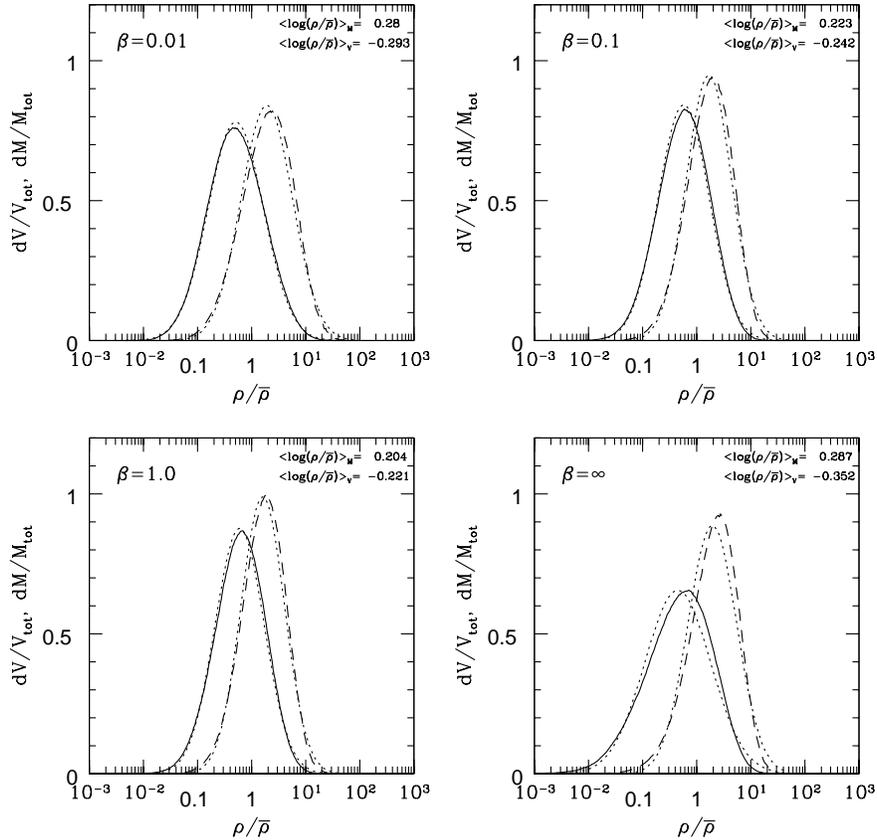}
\end{center}
\caption[]{Distributions of fractional volume (solid curves) and fractional 
mass (dashed curves) as a function of density for Mach-5 forced turbulence 
simulations with varying mean magnetic field strength characterized 
by $\beta\equiv \cs^2/\vA^2$, for $\vA=B_0/\sqrt{4\pi \bar\rho}$. 
Dotted curves show lognormal distributions for comparison. Each panel 
is also labeled with the mass- and volume-weighted mean compression 
magnitudes.}
\label{eps1}
\end{figure}

As the example in Fig. 1 shows, the average mass compression factor is 
relatively insensitive to the mean magnetic field strength.  Although the
minimum value of the mass-weighted mean $\langle \log(\rho/\bar\rho)\rangle_M$
increases (logarithmically) with the fast-magnetosonic Mach number 
${\cal M}_{\rm F}\equiv \vrms/(\cs^2 + \vA^2)^{1/2}$, 
the scatter from ``cosmic variance'' is 
large enough that there does not appear to be a unique relation between 
${\cal M}_{\rm F}$ (or ${\cal M}$; cf. \cite{pad97}, \cite{nor99}) and 
$\langle \log(\rho/\bar\rho)\rangle$  \cite{ost01}.   Typical values of
the mean compression factor for matter in the simulations, 
$\langle \rho/\bar \rho\rangle_M$,
range over $3$--$6$ for ${\cal M}= 5-9$, consistent with 
the compression factor needed to excite the CO molecule  -- rendering 
molecular clouds observable -- when the 
volume-weighted density is only $n\sim 100\, \cm^{-3}$ (e.g. \cite{sco87}). 

More directly observable than the distribution of volume densities in a cloud 
is its distribution of column densities, corresponding to line-of-sight
integrations of density, $\Sigma \equiv \int \rho \D s$.  
Heuristically, one might expect the column density at any projected position 
to be determined by a 
series of independent (in space and time) compressions and rarefactions, 
similarly to the process of events described in (1).  The difference for 
column density is that each event affects only a fraction $f_i<1$ of the 
line of sight, so that the column density is given by 
\begin{equation}
\log(\Sigma/\bar\Sigma)=\uSigma_i \log(1+f_i \delta_i)
\end{equation}
instead of (2). 
The factor $f_i$ may be thought of as the ratio of the correlation length
of a given compression/rarefaction event to the overall linear scale of the 
cloud along the line of sight. If the individual enhancement/decrement 
factors are independent random variables, then the 
resultant column density PDF should take on a log-normal form.
Because each $1+ f_i \delta_i$ is closer to unity than $1+\delta_i$, however,
the mean and dispersion of $\log(\Sigma/\bar\Sigma)$ are expected to 
be smaller than the corresponding quantities for $\log(\rho/\bar\rho)$.
These expectations are indeed borne out by analyses of column densities in 
simulations, as shown by \cite{ost01};  distributions show 
a lognormal form (see also Fig. 2), and typical values of the 
mass-weighted
mean column density are $\langle \Sigma/\bar\Sigma \rangle_M= 1.1 -1.4$ 
for ${\cal M}= 5-9$.

\begin{figure}[t]
\begin{center}
\includegraphics[width=\textwidth]{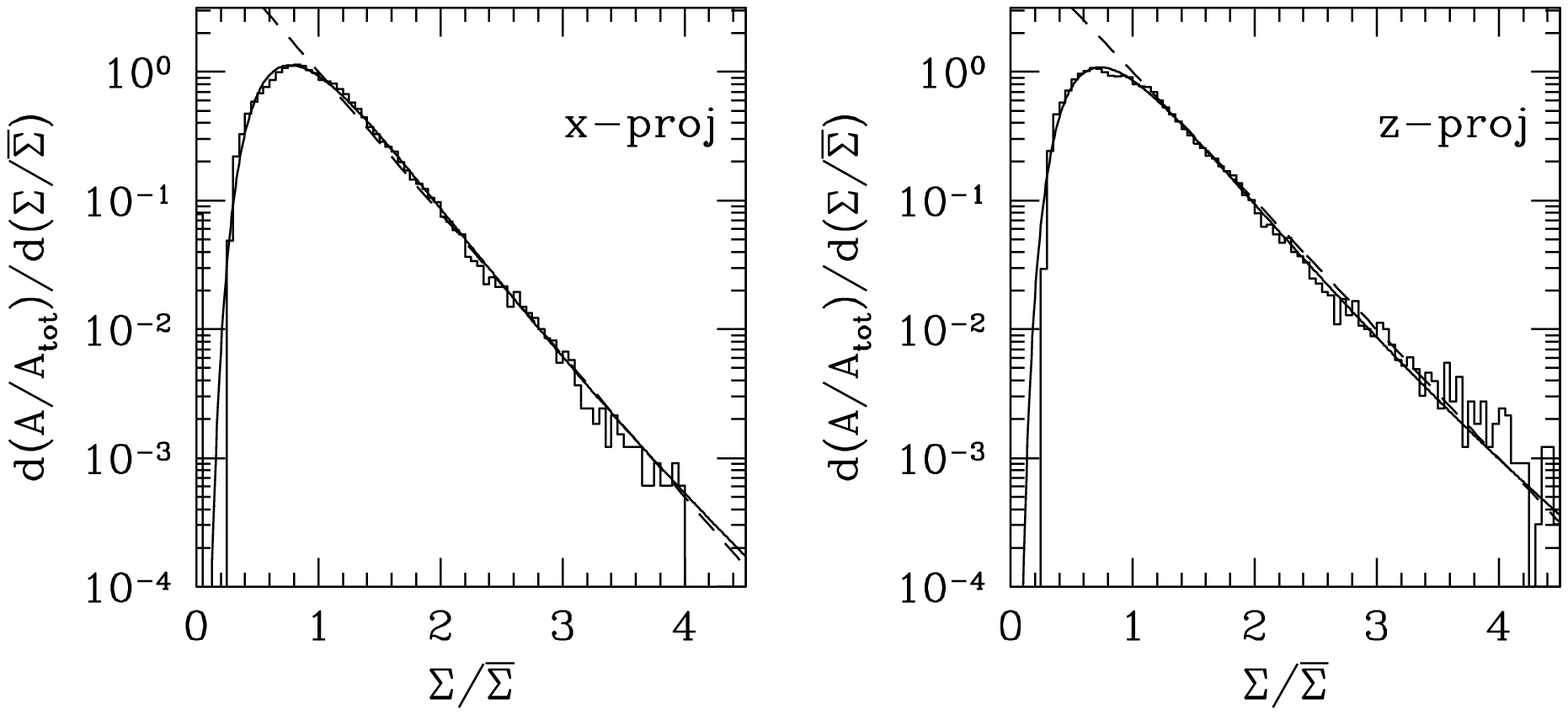}
\end{center}
\vskip -2.5in
\caption[]{Distributions of fractional area (histograms) as a function of
column density for Mach-7 decaying turbulence, for two different projection
directions, from $\beta=0.01$ simulation (see \cite{ost01} for details).   
Solid curves show the
corresponding lognormal distribution fits; dashed lines indicate exponential 
fits on high-column side, with slopes -1.1 and -1 for the $\hat x$ and 
$\hat z$ projections.}
\label{eps2}
\end{figure}

From (3), note that if the factors $f_i$ are small (corresponding to 
having the dominant correlation length small compared to the size of the
numerical box/physical cloud), then $\delta \Sigma/\bar\Sigma\equiv
\Sigma/\bar\Sigma-1\approx\uSigma_i f_i\delta_i$, implying a Gaussian distribution
for  $\delta \Sigma/\bar\Sigma$ if the terms $f_i\delta_i$ are independent
random variables.  Numerical evidence on how column PDFs transition 
from Gaussian to lognormal form as $f_i$ increases is presented in    
\cite{vaz01}.  

Preliminary comparisons between observed PDFs of column density in
filamentary molecular clouds -- obtained from stellar extinction data 
\cite{alv98} -- and simulated PDFs from turbulence models are
very encouraging \cite{ost01}.  The data are consistent with lognormal 
distributions, with comparable width to those from simulations having 
turbulent Mach 
numbers and power spectra comparable to those in observed clouds.  It 
remains to be seen how much more specific information about a cloud can be
learned from its column density.
Although there were earlier some hopes that column density PDFs
could help distinguish the mean magnetic field strength in clouds 
\cite{pad99}, models with matched Mach numbers ``observed'' from varying 
directions do not show strong or consistent correlations of the mean column 
density contrast with the value of 
$\beta\equiv \cs^2(B_0^2/4\pi\bar\rho)^{-1}$ \cite{ost01}.  In spite of this
insensitivity to the mean magnetic field strength, one might still hope to
constrain the distribution of volume densities from the distribution of 
a cloud's column densities.  The mathematical degeneracy between $\delta_i$ 
(the volume compression/rarefaction increment) and $f_i$ (the spatial 
coherence length of an event) evident in (3) indicates that there is 
no simple inversion method.   Methods that combine the column PDF with 
the spatial (two-point) correlation
function in column density maps (reflecting $f_i$, or more generally, the
shape of the density power spectrum) may however 
be able to lift this degeneracy; this represents an important direction 
for future study.

Because the population of the tails of the density and column density PDFs
may be more affected by intermittency than the population near the peak, 
and because the equation of state may depart from isothermality in more 
overdense, optically thick regions (cf. \cite{sca98},\cite{nor99})
observational 
departures from log-normality are more likely to occur there.  Simulations 
(even with a uniform,  isothermal equation of state) 
show a variety of behaviour in the tail distributions, although PDFs 
that are lognormal over more than three orders of magnitude are common 
from isothermal decaying-turbulence simulations. Fig. 2 shows, for example,
that for column densities less than $\Sigma/\bar\Sigma=4$ (corresponding to 
some 98\% of the projected area of the particular 
simulation), a lognormal function is an excellent approximation.

Over a limited range of column densities above the mean, lognormal 
PDFs can typically be fit by an exponential function as well (i.e. 
$\log[dA/d\Sigma]= C_1+C_2 \Sigma/\bar\Sigma$ for $C_1$ and $C_2$ constants). 
Fig. 2 shows examples of local fits of this kind with slopes 
$C_2= -1.1, -1$. 
Similarities between this local exponential form in simulations and in
molecular-line observations \cite{bli97}, with emphasis on 
potential dependences of the slope on the largest density correlation
scale, have been investigated by \cite{bur01}
(see also \cite{ost01} for related discussion of resolution effects).
It is not yet clear whether there is inherent physical significance
in the local exponential form, or whether it is primarily a convenient
mathematical approximation to a lognormal on the high-column-density 
side of the 
distribution where most of the observable matter is found.

\section{Clumps in Turbulent Clouds}

A longstanding unsolved problem in astrophysics is what determines the stellar
initial mass function (IMF).  Many different physical processes could 
potentially affect the IMF; an abbreviated list includes: (i) turbulent
stresses in a large-scale cloud producing non-self-gravitating clumps with
a range of masses/sizes,
(ii) self-gravity in inhomogeneous clumps/cores leading to  
sub-fragmentation of collapsing condensations, (iii) dynamical instabilities 
in massive disks -- formed by the initial collapse of rotating cores --  
fragmenting them into binary or multiple star/disk systems, (iv) outward 
momentum flux from stellar radiation and/or MHD winds truncating 
accretion onto forming stars from the outer parts of their parent cores.  
The relative importance to the final IMF 
of each of these (and other) processes remains to be 
determined, and major technical challenges are involved in attacking 
any of these questions via direction numerical simulations.  The large range 
of scales (nine orders of magnitude!) involved in going from a cloud to a 
star points to the need for adaptive mesh refinement (AMR) 
schemes in order to follow even
a portion of the overall process.

For the present, we can begin by assessing the clumpy structure produced by
turbulence and self-gravity at moderate scales in uniform-grid 
MHD simulations
of GMCs.  This clumpy structure can be characterized in many ways, and 
applying varying methods is valuable for understanding and illustrating 
different aspects of the formal dynamical problem, as well as for interpreting 
observations.

\begin{figure}[t]
\begin{center}
\includegraphics[width=0.6\textwidth]{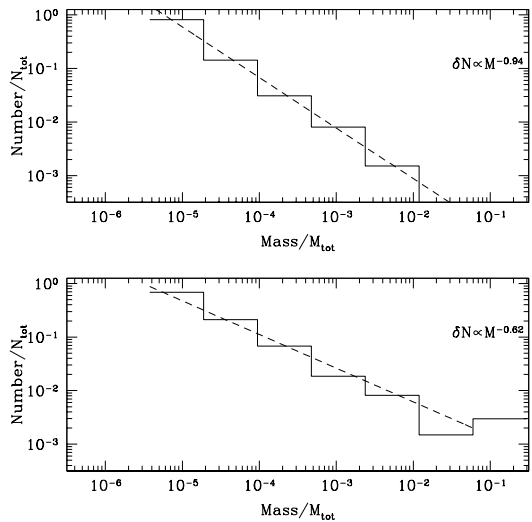}
\end{center}
\caption[]{Number of clumps as a function of mass for two different 
clump identification algorithms. Lower panel shows results of excluding clumps
that lie within larger clumps; upper panel shows results of 
counting clumps at any given spatial scale regardless of whether they lie
within a larger concentration.}
\label{eps3}
\end{figure}

One class of {\it non-hierarchical} structure-analysis methods is similar to 
the CLUMPFIND algorithm introduced by \cite{wil94}.  In this method 
(see \cite{gam01} for details), one 
chooses a threshold volume density $\rho_{\rm th}$ (or column density 
$\Sigma_{\rm th}$), identifies the
set of local maxima in the data cube (potentially first
smoothing the data to reduce pixel noise), and then defines clumps by assigning
matter at $\rho>\rho_{\rm th}$ to the nearest local maximum.  The shape of
a clump can be quantified by computing the ratios of principal axes in
its moment of inertia tensor, and the importance
of self-gravity in binding a condensation can be described, e.g., by 
computing the ratio of gravitational energy to the (weighted) sum of kinetic, 
thermal, and magnetic energies.  

A full discussion of the results of applying this clump-finding method to
a set of decaying-turbulence (with initial ${\cal M}=14$), 
self-gravitating MHD simulations is given in
\cite{gam01}.  Results from this analysis (taking $\rho_{\rm th}/\bar\rho=10$) include:
\begin{itemize}
\item
At any time, only the high-mass wing of the clump distribution is 
self-gravitating;
\item
Characterizing the spectral shape of the high-mass wing as 
$\D N/\D M \propto M^{-x}$, the slope $x$ is in the range 2--3, becoming 
shallower over time from mergers;
\item 
The turnover in the mass spectrum 
is spatially well-resolved (at $\sim 8$ grid zones across) with 
$M_{\rm peak}/M_{\rm tot}$ of a few times $10^{-4}$;
\item 
The minimum clump mass is typically a factor 10 below the peak value;
\item 
Clump shapes are intrinsically triaxial, and project to having two-dimensional
axis ratios $\sim 2:1$;
\item
The shapes of the mass functions of apparent clumps in column density maps 
are similar to those of true three-dimensional clumps, but shifted 
to larger masses by an order of magnitude.
\end{itemize}
An interesting point is that $M_{\rm peak}$, and also the minimum mass 
$M_{\rm min}$ 
of a clump for given $\rho_{\rm th}/\bar\rho=10$,  do not vary 
with the value of $\beta$.  Also, although the Mach number declines by a factor
4--5 over the course of the simulations, the peak and minimum masses do 
not change significantly.  This suggests that the mass function of clumps
retains a ``memory'' of the dynamical history of a cloud, rather than 
being determined solely by the cloud's instantaneous turbulent Mach number and 
spectrum (as proposed in \cite{pad00}).  

The results for typical slopes of the high-mass end of the clump mass function
are intriguingly similar to the value $2.35$ for the Salpeter stellar IMF, 
which also appears to describe the core IMF in forming clusters 
\cite{and00}. 
Similar results have also been obtained by analysis of a variety of 
simulations by other groups (e.g. 
\cite{kle01}, \cite{pnrg00},\cite{bal01}).  While the conclusions from 
these preliminary analyses are promising, many questions still remain open.
It not yet clear in general 
how the mass scales (the peak, minimum, and maximum) depend
on the input parameters ($\cal M$, as well as the spectral shape, and 
potentially $\beta$) and on dynamical history.  Uncertainties also remain
in how three-dimensional clumps relate to clumps seen in projection, and 
whether the latter distribution can be used to deduce the former.  

Another important set of questions is how the definition of a ``clump'' --
both the specific clump identification algorithm with its chosen set of 
parameters (such as $\rho_{\rm th}$), and the overall category of algorithms
in which a specific method lies -- affects the results.  A clear categorical 
distinction is between  
non-hierarchical identification algorithms like 
CLUMPFIND (in which every mass element is assigned to a single clump), and
hierarchical algorithms, in which a given mass element may be counted as part
of many clumps, at different levels of a hierarchy.  As an astronomical 
analogy, it is clear that for many purposes it 
is valuable to count galaxies whether or not they are 
part of larger clusters or supercluster; counting ``objects'' is a function 
of the spatial scale under consideration.

As an example of 
how hierarchical considerations affect the mass spectrum, consider the 
distributions shown in Fig. 3.  For this analysis,  a ``clump'' is 
any cubic region at a given spatial scale in which the density exceeds the 
mass-weighted mean density for the ensemble of cubes at that scale 
(see \cite{ost01} for further details on this ``region of contrast'' 
algorithm).  For the lower panel in Fig. 3, only clumps that do not 
lie within other clumps are counted for the mass spectrum;  for the upper
panel, the spectrum counts clumps regardless of ``overlap.''  As should
be expected, allowing for clumps-within-clumps leads to a relatively 
steeper mass spectrum, with a slope $\D N/\D M\propto M^{-1.9}$ in this 
instance.  The no-overlap spectrum is $\D N/\D M\propto M^{-1.6}$.  This
difference between the hierarchical and non-hierarchical mass spectra is
very interesting because it is reminiscent of the difference between the 
steep stellar IMF, and shallower observationally-determined 
GMC mass functions (e.g. \cite{sol87}).
Because the density condensations produced by turbulence as ``initial 
conditions'' for collapse are hierarchically
nested, this difference in slopes offers intriguing support for 
the idea 
that fragmentation during gravitationally-collapsing stages may play 
a crucial role in defining the stellar IMF.  An important question for 
future AMR simulations to address is when and why gravity chooses an 
``inner'' versus ``outer'' mass scale for final collapsed objects.

\section{Linewidth-size Relationships}

The direct observables produced by spectral-line mapping of a molecular cloud
are data cubes of intensity as a function of two plane-of-sky positions and
the line-of-sight velocity (in radio astronomy, the intensity is described 
as a brightness temperature).  In principal, one would like to extract the
spatial distribution of velocity and emissivity from these data cubes.  Because
only projected data are available, and a turbulent cloud has no spatial 
symmetries to exploit, direct inversion is not possible.  However, one may 
still hope to deduce statistical properties of the turbulence from the
full intensity data cube.  Various complex techniques to do this are under 
development by several groups -- including Principal Component 
Analysis \cite{hey97}, \cite{bru01}; the Spectral Correlation Function
\cite{ros99},\cite{pad01}; and Velocity Channel Analysis \cite{laz00},
\cite{laz01}.  Here, I will briefly discuss a simple technique 
to estimate a theoretically fundamental -- and observationally 
much-investigated -- property of turbulence, the variation of linewidth
with physical size scale.  

Averaged over volumes with the {\it same size} in all three directions,
the mean linewidth simply reflects the 
underlying one-dimensional velocity power spectrum, since 
\begin{equation}
\sigma_v^2(s)=\int\int\int_0^s \D^3x\, v^2(\vec{x})
={1\over (2\pi)^3}\int\int\int_{2\pi/s}^\infty \D^3k\, v^2(\vec{k}). 
\end{equation}
If the turbulence has a power-law spectrum, 
$v^2(\vec{k})\propto |k|^{-\alpha}$, then 
$\sigma_v(s)\propto s^{(\alpha-3)/2}$, so for e.g. Kolmogorov or Burgers 
spectra with $\alpha=11/3$ or $4$, $\sigma_v\propto$ $s^{1/3}$ or $s^{1/2}$.

In observations, however, any region of size $s^2$ in projection (on
the plane of the sky) in general extends over a scale at least as large  
along the line of sight.  If $s$ is small compared to the overall scale of
a cloud, $L$, then the observed linewidth from a region of projected area $s^2$
can have contributions from $L/s\gg 1$ volume elements of size $s^3$ 
along the line of sight.  If power increases with scale ($\alpha>3$), then
the velocity centroids of the multiple $s^3$ volume elements {\it on average} 
differ, such that the linewidth over the area $s^2$ 
integrated over the whole line of sight will {\it on average} 
exceed $\sigma_v(s)$, reaching up to $\sigma_v(L)$. However, because of the
statistical nature of the distribution (e.g. if it obeys Gaussian random 
statistics), for {\it some} projected positions the centroids of the 
multiple $s^3$ volume elements will differ very little, such that the 
line-of-sight integrated velocity dispersion will be close to 
$\sigma_v(s)$.  From this argument, one expects that the {\it mean linewidth} 
would vary weakly with projected size, whereas the {\it lower envelope}
of the linewidth distribution would vary more strongly with projected size,
and in fact trace the underlying three-dimensional mean linewidth-size 
relation.  Analysis of simulation data cubes indeed bears out this 
expectation \cite{ost01}, as shown for example in Fig. 4.  
\begin{figure}[t]
\begin{center}
\includegraphics[width=0.6\textwidth]{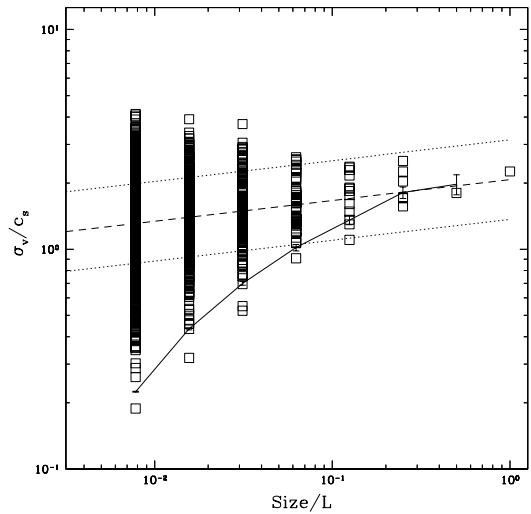}
\end{center}
\caption[]{Linewith-size relations for projected and 3D simulated data.
Distribution (squares) shows total linewidth for 
apparent clumps in a map vs. linear dimension of projected clump 
(i.e. square root of its area); dashed line 
shows (linear) fitted mean linewidth-size distribution for projected clump map
(slope is 0.2).  Solid curve shows mean linewidth-size relation for 
3D volumes from the same data cube.  Data set is $\beta=0.1$ model with 
Mach number 5 from \cite{ost01}, with threshold density $\rho/\bar\rho=3$.
}
\label{eps4}
\end{figure}

The foregoing discussion is helpful for interpreting well-known 
observational aspects of molecular cloud kinematic scalings.  
In particular, by relation to Fig. 4, one may 
understand why the mean linewidth-size relation for apparent clumps 
observed in molecular tracers with 
moderate $\rho_{\rm crit}/\bar\rho$ is relatively flat (e.g. 
\cite{stu90},\cite{ber92},\cite{wil94}) 
compared to the relatively steeper ``Larson's Law'' 
($\sigma_v\propto s^{1/2}$) linewidth-size relations 
(cf. \cite{lar81}, \cite{sol87}) that apply to objects
that are spatially ``isolated'' -- either because 
$\rho_{\rm crit}/\bar\rho$
is large (for dense cores within clouds) or because of phase differences 
with their surroundings (for molecular clouds within the atomic ISM).  
For tracers with critical density near the (mass-weighted) mean density in
a cloud, it is not unlikely for multiple structures that are separated
and in relative motion 
along the line of sight to contribute to the linewidth at a given 
position on a map.  For tracers with higher critical densities, the 
probability of chance projections of dense condensations is much lower because
there are many fewer such condensations.
Spectral correlation function analyses \cite{pad01} also demonstrate that 
simulated spectral data cubes from higher-density tracers yield more spatial 
variation than do data cubes from lower-density tracers, 
for the same reason: the spectrum at a given point in a map
from a low-density tracer samples 
more completely along a line of sight, and is thus more representative 
of the mean spectrum averaged over a whole cloud -- compared to the spectrum
from a high-density tracer.

This discussion of linewidth-size relations in 2D and 3D 
also serves to illustrate the point that coherent 
structures in position-velocity space in general differ from 
coherent structures in 3D position (physical) space;  
detailed analyses of simulation data cubes have shown this in a number of 
different ways (e.g. \cite{ost01},\cite{pic00},\cite{bal01}).  As a 
consequence, mass functions of density condensations, or other 
measures of structure in the density such as its Fourier power spectrum, 
cannot necessarily be obtained by treating the line-of-sight velocity as a 
surrogate for 
line-of-sight position.  Instead, it is necessary to use statistical 
approaches to diagnose structure in the physical density distribution -- 
just as statistical approaches are required for diagnosing structure in the 
velocity distribution.  Depending on the relative power in velocity and 
density fluctuations, true density structure may be best discerned by 
integrating intensity over velocity and then correcting for physical 
superposition of dense structures using information from two-point 
correlation functions in the column density (see \cite{laz00} for related
discussion).

\section{Polarization as a Magnetic Field Diagnostic}

It has long been recognized that polarization studies are important for
diagnosing basic properties and structure of the ISM, because they provide 
relatively direct access to the elusive -- but dynamically important -- 
magnetic field (e.g. \cite{hei96},\cite{zwe96}).  {\it Provided} that dust 
grains preferentially
align with short axes parallel to the local direction of the 
magnetic field (\cite{dav51}; see e.g. recent review of 
\cite{lazar00} for thorough discussion of alignment mechanisms), 
an ordered $\vec{B}$-field will lead to observable 
polarization.  For grains aligned with their short axes parallel to the 
local magnetic field, background stars are observed in optical/near-IR 
wavelengths with polarization parallel to the local magnetic field, and 
local dust emission is observed in far-IR/sub-mm wavelengths with polarization 
perpendicular to the 
local magnetic field.  If the angle of $\vec{B}$ with respect to the
plane of the sky is $i$, the local contribution to polarization is
$\propto \cos^2 i$ times the difference between long- and short-axis 
grain crossections times the density of polarizing grains.  Taking the 
simplest-possible assumption of a uniform ratio of polarizing-grain density 
to gas density (but see below), it is straightforward to create simulated 
polarization maps by integrating the radiative transfer equations for 
the Stokes parameters (\cite{mar74},\cite{mar75},\cite{lee85}).  For 
a medium with optical depth $\tau\ll 1$, the fractional polarization in 
(thermal) emission and in dust-absorbed starlight are related by 
$P_{\rm em}=P_{\rm abs}/\tau$ (e.g. \cite{hil88}), so that polarization
from emission is proportional
to polarization from extinction divided by the column of intervening matter.

\begin{figure}[t]
\begin{center}
\includegraphics[width=1.\textwidth]{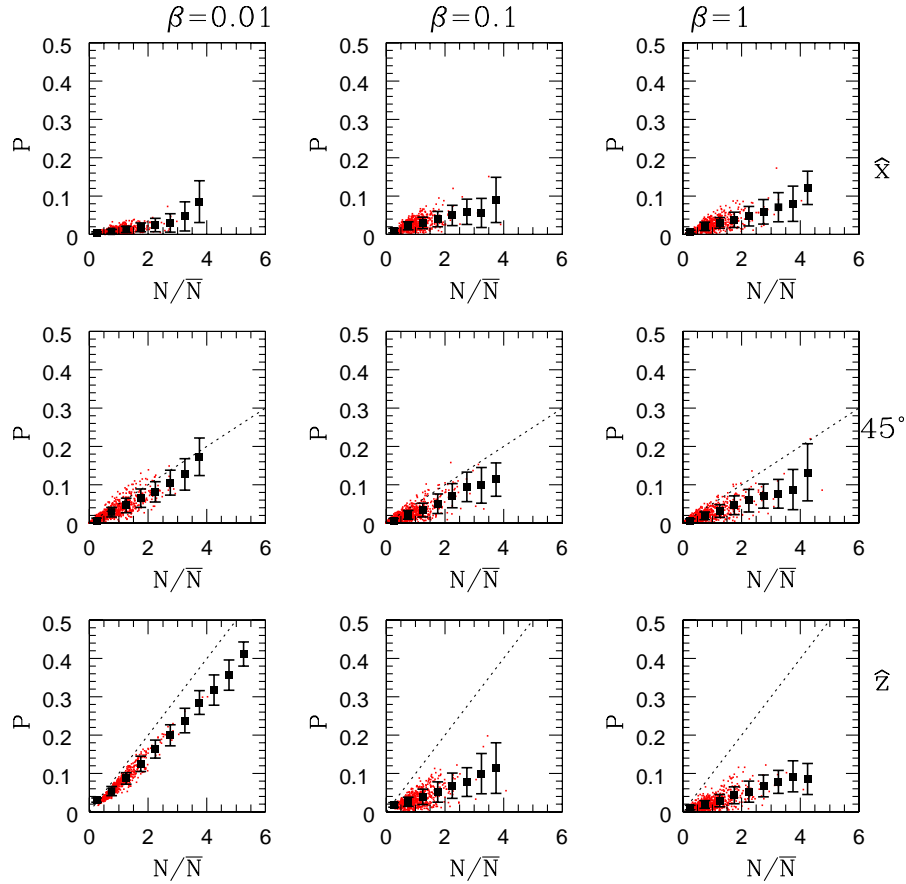}
\end{center}
\caption[]{Distributions of simulated polarized-extinction as a function 
of column density, for decaying-turbulence simulations with $\cal M$=7 from
\cite{ost01}.  Scatter plots show $P_{\rm abs}$ for a randomly-chosen 
subset of 
positions on the projected map, for three different projection directions, and
three different mean magnetic field strengths.  Squares with 1-$\sigma$ error 
bars show the mean $P-N$ relation for the full data sets. 
Dotted lines show what the $P-N$ relation would be for uniform magnetic fields.
$P$ is arbitrarily normalized to a value 
0.1 for uniform density $N=\bar N$ for a projection perpendicular to 
$\hat B_0$}
\label{eps5}
\end{figure}

Maps and analyses of polarized extinction \cite{ost01}, 
and emission \cite{pgdjnr01},\cite{hei01} computed from
MHD simulations have recently been presented by several groups.  One 
question of interest is how the fractional polarization varies with the
column $N\equiv \int n \D s$ of absorbing or emitting material.  
For a uniform magnetic field,
uniform polarization efficiency of grains, but spatially-nonuniform 
distribution of matter, $P_{\rm abs}$ would increase linearly with $N$, 
while $P_{\rm em}$ would be independent of $N$.  For a spatially nonuniform
field, the situation is much more complicated.  In this case, the increase in
$P_{\rm abs}$ or $P_{\rm em}/\tau$ with $N$ would (a) in general be shallower 
because variations in the magnetic field direction 
decorrelate the direction of grains, reducing the net contribution to 
polarization per unit length along the line of sight, and (b) no longer follow
a linear relation if the amplitude of field fluctuations is large and/or 
if the number of effective correlation lengths of 
magnetic field orientation along the line of sight varies with 
column density (this can yield, e.g., $P_{\rm abs}\propto N^{1/2}$ 
\cite{mar74}).
For weaker mean magnetic fields $B_0$ and a given power spectrum 
of fluctuations 
$\delta B$, the directional decorrelation in polarization occurs at a 
physically smaller scale than for the strong-$B_0$ case, so that lower 
polarization is expected for a given column density.  These expected trends
are indeed evident in distributions of $P_{\rm abs}$ vs. column density
from MHD simulations, as shown for example in Fig. 5.

Interestingly, observed distributions (e.g.\cite{goo95}, \cite{arc98}) of
polarized-extinction vs. column (or $A_V$) in molecular clouds do not
show the behavior evident in Fig. 5, which would correspond to a
secular increase in $P_{\rm abs}$ up to $A_V$ of 30 or more (for $\bar
N$ corresponding to a typical GMC $A_V\sim 7.5$ \cite{mck99}).
Instead, the observed increase of $P_{\rm abs}$ with $A_V$ flattens
near $A_V$ of unity, possibly indicating that grain alignment fails in
the deep interiors of clouds (e.g. \cite{laz97}).   Additional support for this
idea comes from comparison of observed distributions (e.g. \cite{hen01}) of 
$P_{\rm em}$ vs. intensity (proportional to $N$) in dense cores 
with simulated distributions; the simulations show insufficient 
decrease in $P_{\rm em}$ with column unless high-$A_V$ regions have 
decreased polarizing efficiency \cite{pgdjnr01}.  

Potentially, one of the most important applications for 
polarization studies is to use the variation in polarization directions to
diagnose the strength of the magnetic field \cite{cha53}.  The basic 
physical idea behind the so-called ``Chandrasekhar-Fermi'' method is that 
weaker magnetic fields produce lower tension forces for a given displacement, 
so that for a given velocity field, a lower mean magnetic field strength 
results in larger distortions in the magnetic field direction.  For the
case of a single low-amplitude Alfv\'en wave in a uniform magnetic field with 
plane-of-sky component $B_p$, the magnetic field and velocity perturbations 
obey $\delta B/B_p=\delta v/v_{\rm A, p}$, where 
$v_{\rm A, p}=B_p(4\pi \bar\rho)^{-1/2}$.  For a wave with 
perturbation direction in the plane of the sky, the dispersion in
polarization directions is $\delta \phi= \langle (\delta B/B_p)^2\rangle^{1/2}$
if the net polarization is either parallel or perpendicular to $\vec{B}$.
If there is an Alfv\'en wave component with perturbations along the 
line of sight having the same amplitude as the component with perturbations in
the plane of the sky, then 
$\delta v /v_{\rm A, p} \equiv 
\langle v_{\rm los}^2\rangle^{1/2}/v_{\rm A, p}
= \delta \phi$, so that $B_p$ is 
given by 
$B_{\rm CF}\equiv (4 \pi \bar\rho)^{1/2} \delta v /\delta\phi$.  
Thus, with measures of
the mean density, the observed velocity dispersion, and the dispersion in
polarization angles, the plane-of-sky field strength may in principle be 
estimated.

\begin{figure}[t]
\begin{center}
\includegraphics[width=0.6\textwidth]{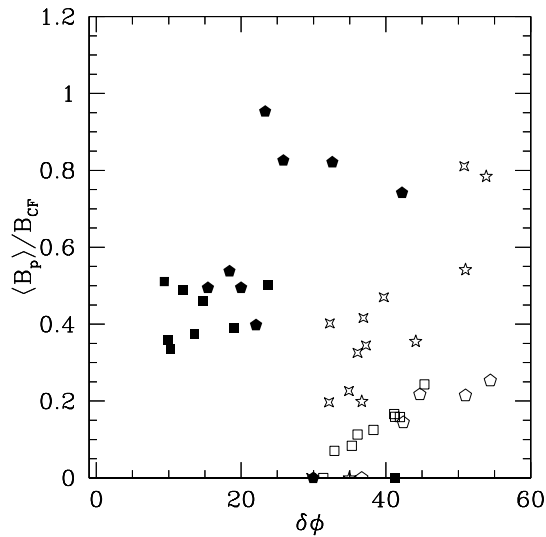}
\end{center}
\caption[]{Plane-of-sky component of the mean magnetic field, $B_p$, 
relative to Chandrasekhar-Fermi estimate $B_{CF}$, 
for $\cal M$=7 (four-sided symbols)
and $\cal M$=9 (five-sided symbols) decaying turbulence MHD simulations 
\cite{ost01}; $\delta \phi$ is the corresponding dispersion (in degrees)
of directions in the polarized-extinction map.
Solid, starred, and open symbols represent $\beta=0.01,\ 0.1,\ 1.0$ models.
}
\label{eps6}
\end{figure}

Since several of the idealizations described above do not hold in
a real molecular cloud, it is desirable to test and/or recalibrate the 
Chandrasekhar-Fermi (hereafter ``C-F'') 
relation using MHD simulations, which allow for more
complex dynamical structure.  Figure 6 shows an example of such a test 
using simulated polarized-extinction data, comparing the true mean 
plane-of-sky 
field strength with the ``one-wave/equipartition'' C-F
estimate given above.  As reported in \cite{ost01}, when the dispersion in 
polarization angles is sufficiently small ($\delta \phi<25^\circ$) -- 
such that linear theory is adequate, a good estimate
of the plane-of-sky field is $\langle B_p\rangle \sim 0.5 B_{CF}$.   
Evidently, the 
presence of more than one wave along the line of sight reduces $\delta \phi$,
so that $B_{CF}$ tends to overestimate the true $B_p$.
For larger dispersions in polarization angle, Fig. 6 shows that 
the measure $0.5 B_{CF}$ ``calibrated'' at small $\delta \phi$  may either 
under- or over-estimate $B_p$.  Note that even if the mean magnetic field 
is strong, $\delta \phi$ can be large -- and linear theory inappropriate -- 
for system orientations in which the mean magnetic field direction 
lies near the 
line of sight.   Analyses of simulated polarized-emission maps, combined with 
kinematic measurements \cite{hei01}, or with synthetic radiative transfer
spectral maps \cite{pgdjnr01}, yield similar results for testing of the C-F
formula.

\begin{figure}[t]
\begin{center}
\includegraphics[width=0.6\textwidth]{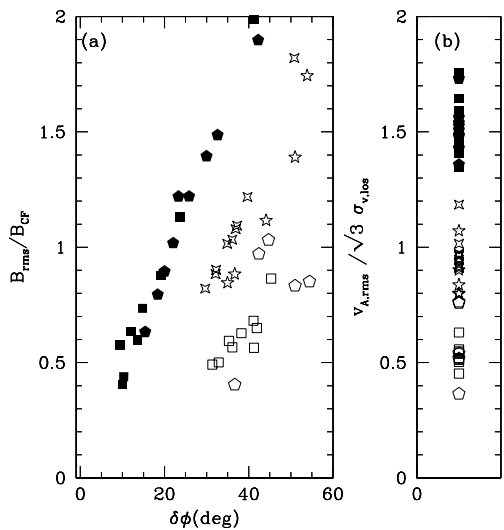}
\end{center}
\caption[]{(a) Total (rms) magnetic field strength 
relative to C-F estimate $B_{CF}$ vs. 
dispersion (in degrees) of directions in the polarized-extinction map.
 (b) Total (rms) Alfv\'en speed compared
to $\sqrt 3$ times the line-of-sight velocity dispersion.
Data and symbols for (a) and (b) are as in Fig. 6.
}
\label{eps7}
\end{figure}

We note that the C-F
method measures only the mean 
{\it plane-of-sky} magnetic field $\langle B_p\rangle $; Figure 7(a) 
shows that even for small dispersion in polarization angles, there can be 
large variation in the {\it total} field strength compared to the ``corrected''
C-F estimate.  In any individual cloud, the line-of-sight field can 
be estimated by the Zeeman effect (at least in principle; in practice this is 
difficult), with the two estimates combined to obtain the 
total three-dimensional field strength (e.g. \cite{mye91}).
Another potential caveat for observational application of 
the C-F method is that the principal correlation scale for the
dispersion in magnetic field directions must be resolved in the plane of
sky by the effective observational ``beam''.  For polarized extinction, 
this poses no difficulties because background star observations have
minimal beam thickness.   For polarized emission from warm 
cores, however, large-beam averaging of spatial fluctuations tends to 
reduce $\delta\phi$ relative to its value for a ``pencil-beam'' observation,
potentially resulting in an overestimate of $B_p$ 
unless an appropriate correction factor is applied \cite{hei01}.    

When variations in the polarization directions in a map are large, the 
fluctuations of the plane-of-sky magnetic field must be comparable to the 
mean value of $B_p$.  In this case, it is not possible to distinguish
from polarization studies alone
whether (a) a cloud has a large mean magnetic field that is ``hidden'' along 
the line of sight (as is the case for the solid points in Fig. 6 having 
large $\delta\phi$), or (b) the mean magnetic field is simply weak 
(as is the case for the starred and open points in Fig. 6).  Observed 
line-of-sight velocities combined with an assumption of 
equipartition between magnetic and kinetic energies can still yield 
an approximate measure of the rms magnetic field strength (see e.g. 
Fig. 7b) that is correct within a factor $\sim 2$, 
although an important caveat is that this estimate could be 
arbitrarily far off for cases with very strong mean magnetic fields 
($\vA\gg v_{rms}$) that happen to lie near the line of sight.
\footnote{Also note that while $B_{rms}$ is a dynamically-important
quantity, it is not equivalent to the mean magnetic field strength 
$|\langle \vec{B}\rangle|$ that enters into the mass-to-flux ratio, which ultimately
determines whether a cloud or core is super- or sub-critical.}  Fitting 
formulae that interpolate between the small-$\delta\phi$ and 
large-$\delta\phi$ limits (with the implicit assumption that $\vA$ along
the line of sight is not very large)
have been provided by \cite{hei01}.  

\vfil
\section{Summary}

To understand the intrinsic nature of MHD turbulence and the relation between
the turbulence observed in astronomical systems and the turbulence 
simulated numerically, it is crucial to develop structural diagnostics.  
These diagnostics may:
\begin{itemize}
\item
Enhance conceptual understanding of the turbulent phenomenon;
\item
Enable determination of astronomical systems' properties that are either
difficult to observe at all (e.g. $\vec{B}$) or only indirectly 
observable because of projection effects (e.g. $\vec v_k$, $\rho_k$);  
\item
Provide a physical basis or interpretation for observed empirical ``laws''
(e.g. column density distributions, linewidth-size relations, 
mass functions of clouds, clumps, and stars);
\item 
Indentify when additional physical ingredients may be needed in a 
computational model.
\end{itemize}
From the examples outlined in this paper, it is clear that significant 
advances along these lines have already been accomplished.

Conceptual and (approximate) quantitative understanding of what determines 
the PDFs of density and column density observed in 
molecular clouds have already been obtained from analyses to date (see \S 2).
The development of lognormal statistics from Gaussian random processes 
under near-isothermal conditions appears to obtain robustly for a variety
of conditions, with little sensitivity to the magnetic field strength.
Because one-point density statistics are subject to 
``cosmic variance'' (if low wavenumbers dominate the power spectrum,
different realizations of a given power spectrum have significant variation),
and because degeneracies prevent the direct inversion of column density 
statistics to obtain volume density statistics,
there may not be a highly accurate way to determine the physically-important 
volume-averaged density in a cloud from more direct observables
such as the observed velocity dispersion and observed  distribution of 
extinction.  Future work that combines one-point column density 
statistics with two-point correlation functions could potentially be valuable 
in constraining the spatial power spectrum of density fluctuations.  
Such analyses could also be useful in relating mass functions
of clumps seen in projection with true three-dimensional clumps (see 
\S 3, \S 4).

Analysis of clumpy structure in model clouds (see \S 3) shows intriguing 
correspondence to observations: mass functions of 
self-gravitating structures are comparable to the Salpeter IMF, and 
there are significant differences between steeper (more 
``stellar-like'') mass functions when subcondensations are not subsumed into 
larger structures, and shallower (more ``cloud-like'') mass functions 
when substructure is discounted.  
These findings support the concepts that
clumping imposed by turbulence, as well as fragmentation during 
gravitational collapse, may both be important in establishing the IMF.
Future work is required to determine 
whether there might be a relationship between Mach numbers (instantaneous 
and historical), the overall size and mass of a cloud, and  
the characteristic ``peak'' sizes and masses of self-gravitating condensations 
within a simulation.  
More coverage of model parameter space (allowing 
for different power spectra, forcing, etc.) will be important
for deciding how sensitive the resulting clump mass functions may be to 
physical conditions.

Some of the longest-established empirical results about molecular clouds 
concern the correlations among physical scale and spectral linewidth, and 
analyses of simulations have proven valuable in interpreting how these 
empirical ``laws'' relate to the underlying properties of turbulent 
clouds and how they are observed (see \S 4).  Steep (``Larson's law'') 
dependence of linewidth on size probably reflects the true three-dimensional 
power spectrum, since the structures to which these steep laws apply
are observed in tracers that render them well-separated from the background.
On the other hand, the weak dependence of mean linewidth 
on 2D size of apparent moderate-density clumps 
within clouds may largely arise from projection effects -- with the 
scale sampled by the velocity dispersion on average extending over 
much of the whole cloud's depth.  Both simple methods using the lower envelope 
of the linewidth-size distribution -- and more complicated methods using 
detailed spectral shapes and their spatial correlations, 
are very promising for being able to distinguish the true three-dimensional
power spectrum from molecular line data cubes.

Since magnetic fields are difficult to measure directly, there are particularly
strong motivations to develop indirect diagnostics (see \S 5).  
Polarization either 
in absorption or emission is sensitive to the local direction of the magnetic
field, and the variation in the local direction of the magnetic field 
is sensitive to the magnitude of the magnetic field and level of turbulence
in a cloud.  Thus, one might hope to combine observed measures of 
variation in the polarization direction with observed measures of turbulence 
via  molecular linewidths to infer the mean magnetic field strength; 
this kind of indirect method was originally proposed by Chandrasekhar and 
Fermi.  Testing the Chandrasekhar-Fermi method with simulation ``data'' 
shows that for low dispersion in the polarization angle, recalibration
by a factor one-half indeed yields a good measure of the mean plane-of-sky
magnetic field.  When the polarization direction has large
fluctuations, the Chandrasekhar-Fermi method loses accuracy, but simulations
show that an assumption of magnetic/kinetic equipartition is usually 
correct within a factor $\sim 2$.  
Because underresolution tends to enhance the estimated
field strength, and because high-column/high-density regions may not be 
efficient polarizers, there are some potential caveats in applying the
Chandrasekhar-Fermi method
for polarized-emission data.  The promise shown by analyses to date, together
with the potential to obtain large-scale maps of polarized absorption and 
emission in molecular clouds, marks this area as an important direction 
for further research.

These detailed results are exciting, and represent only a small sample of 
the progress that has taken place in this field to date.  
Perhaps the most fundamental advance, however, has been the movement to
match our sophisticated concept of
what a molecular cloud {\it is} -- a complex structure with multiple-scale,
large-amplitude, turbulent fluctuations in all fluid variables; 
with a commensurately sophisticated way to diagnose structure --
by developing and testing analytical tools on detailed 
simulation data cubes that are self-consistent, time-dependent solutions
of the MHD equations.  Recent accomplishments have greatly advanced 
our field, but we are still very much in the era of discovery. Ongoing
interplay between simulation, analysis, and observation will be essential 
for continued 
progress in building a comprehensive dynamical model of molecular clouds.


\bigskip\bigskip

I am grateful to J. Stone and C. Gammie for permission to present results 
from collaborative work, and to the referee A. Lazarian for helpful 
comments.  This research is supported by NASA grants 
NAG 53840 and NAG 59167.

%

\end{document}